\def \be {\begin{equation}}
\def \eq {\end{equation}}
\def \bee {\begin{eqnarray}}
\def \eqq {\end{eqnarray}}
\def \nn {\nonumber}
\def \bea {\begin{array}{c}}
\def \eqa {\end{array}}

\def \la {\langle}
\def \ra {\rangle}
\def \R {{\bf R}}
\def \C {{\bf C}}
\def \Z {{\bf Z}}
\def \del {\partial}
\def \dels {\partial\kern-.5em / \kern.5em}
\def \As {{A\kern-.5em / \kern.5em}}
\def \Ds {D\kern-.7em / \kern.5em}

\def \a {\alpha}
\def \b {\beta}
\def \dag {\dagger}
\def \g {\gamma}
\def \G {\Gamma}
\def \d {\delta}
\def \eps {\epsilon}

\def \Lam {\Lambda}
\def \s {\sigma}
\def \r {\rho}
\def \om {\omega}
\def \Om {\Omega}
\def \one {{\bf 1}}
\def \th {\theta}
\def \Th {\Theta}

\def \II {I\hspace{-.1em}I\hspace{.1em}}

\def \IIB {\mbox{\II B\hspace{.2em}}}

\def \U {{\cal U}}
\def \Ut {\tilde{U}}
\def \H {{\cal H}}
\def \Rt {\tilde{R}}
\def \at {\tilde{a}}
\def \A {{\cal A}}
\def \At {\tilde{A}}
\def \ti {\tilde{i}}
\def \D {{\cal D}}
\def \Dt {\tilde{\cal D}}
\def \pih {\hat{\pi}}
\def \Cc {{\cal C}}

\def \pih {\hat{\pi}}
\def \gh {\hat{g}}
\def \Re {\mbox{Re}}
\def \Im {\mbox{Im}}

\documentclass[12pt]{article}

\begin{document}
\begin{titlepage}
%\catcode`\@=11
%\catcode`\@=12
%\twocolumn[\hsize\textwidth\columnwidth\hsize\csname%
%@twocolumnfalse\endcsname

%\draft
\begin{center}

\hfill hep-th/9812143\\
\hfill APCTP-98-025\\

\vskip .5in

\textbf{\large Matrix Compactification On Orientifolds}

\vskip .5in

Pei-Ming Ho$^1$ and Yong-Shi Wu$^2$

\vskip .3in

\slshape{$^1$ Department of Physics, \\
National Taiwan University, Taipei 10764, Taiwan}\\
\slshape{$^2$ Department of Physics, University of Utah,\\
Salt Lake City, Utah 84112, U.S.A.}\\

\vskip .2in
\sffamily{pmho@phys.ntu.edu.tw \\
wu@mail.physics.utah.edu}

%\maketitle
\end{center}

\vskip .5in

\begin{abstract}

Generalizing previous results for orbifolds, in 
this paper we describe the compactification of 
Matrix model on an orientifold which is
a quotient space $\R^d/\Gamma$
as a Yang-Mills theory
living on a quantum space. The information of the 
compactification is encoded in the action of 
the discrete symmetry group $\Gamma$ on Euclidean 
space $\R^d$ and a projective representation $U$ of $\G$.
The choice of Hilbert space on which the algebra of $U$
is realized as an operator algebra corresponds to the 
choice of a physical background for the compactification.
All these data are summarized in the spectral triple
of the quantum space.

%\pacs{PACS numbers: 11.25.-w, 11.25.Mj, 11.25.Sq}%]

\end{abstract}

\end{titlepage}

%\begin{narrowtext}

\setcounter{footnote}{0}

\section{Introduction} \label{Intro}

The original Matrix model \cite{BFSS} was formulated 
as a microscopic model for the M theory
in eleven dimensional spacetime.
To describe the real world one needs 
to understand how to compactify the theory to lower 
dimensions. In the absence of a rank-three
antisymmetric tensor field background, the description 
of the compactification of Matrix model is similar to 
that of D-branes in compactified string theory
\cite{DM,JM,Tay,GRT}. However, if such a background is 
present, we have to use the concepts of noncommutative 
geometry to describe the compatified Matrix model, as 
first shown by Connes, Douglas and Schwarz \cite{CDS} 
for toroidal compactification. 

In refs.\cite{HWW,HW3} we generalized the noncommutative 
geometric description of D-branes in ref.\cite{HW1}
to incorporate orbifold compactifictions,
including the possibility for discrete torsion.
In particular, in ref.\cite{HW3} we gave a general 
formulation for interpreting the Matrix model compactified 
on manifolds and orbifolds as gauge field theories on 
quantum spaces. In this paper we proceed to describe the 
compactified Matrix model on an orientifold, which is 
the quotient of the flat infinite space by the action of 
a discrete symmetry group $\G$ of the Matrix model, which 
contains the operation of matrix transposition. More 
precisely, we will allow a $\Z_2$-grading of $\G$, which 
corresponds to transposition of the matrix variables
$X^{\mu}$ and $\Psi^{\a}$. If the Matrix model action is 
interpreted as the low energy effective action of D-branes,
the transposition represents the worldsheet parity
transformation of the open strings connecting the D-branes.
In this way, the quotient space is an orientifold 
if $\G$ has such nontrivial $\Z_2$-grading.

Our presentation will be such that the quotient space can 
be a manifold, orbifold or orientifold. This paper also
contains some technical details which were not
included in our previous short letter \cite{HW3}. 
We review how the quotient conditions
arise from taking quotient of the flat space
in Sec.\ref{QuotCond}. We will see that
the quotient conditions need to be supplemented by
the information of the $U$-algebra (\ref{UU1})
in order to completely specify the compactification.
In Sec.\ref{coho} we show that the moduli space of
the $U$-algebra is a $\Z_2$-graded second Hochschild
cohomology group $H^2(\G,U(1))$. For the trivial bundle 
on the dual quantum space, the most general solution 
to the quotient conditions are obtained in Sec.\ref{Sols}.
Following a brief review of noncommutative geometry
in Sec.\ref{NC}, we give the noncommutative geometric
description of the dual quantum space in Sec.\ref{QS}.
Then we examine residual symmetries in the compactified 
Matrix model in Sec.\ref{ResSym}, and give a few examples 
to demonstrate how the formulation actually works in 
Sec.\ref{Example}. Finally in Sec.\ref{General} we 
comment on several possible generalizations of our 
formulation of Matrix model compactifications.

\section{Quotient Conditions} \label{QuotCond}

Consider the compactification of Matrix model
on a space which is the quotient of the flat 
space $\R^d$ over a discrete group $\G$. Taking the 
quotient makes sense for the matrix model only 
if $\G$ is a symmetry of the Matrix model. Let 
the matrix variables be $(X^{\mu}, \Psi^{\a})$,
where $\mu=0,1,\cdots,9$ and $\a=1,2,\cdots,16$
are the Lorentz indices for spin $1$ and spin 
$1/2$ representations, respectively, in $10$ 
dimensional Minkowski space in the light-cone
formulation for M-theory. In addition to the 
usual isometry group $G_0$ of the superspace, 
which is the super-Poincar\'{e} group times 
the spacetime reflections and reversals,
the matrix transposition of these variables is 
also a symmetry of the Matrix model.
In the D0-brane interpretation
for the original Matrix model \cite{BFSS}, the 
matrix index labels different D0-branes, so that 
off-diagonal elements of $X^{\mu}$ and $\Psi^{\a}$ 
describe open strings stretched between the 
D0-branes. Therefore taking transposition means 
string worldsheet parity transformation.
Since $X^{\mu}$ and $\Psi^{\a}$ are Hermitian 
matrices, taking matrix transposition is 
equivalent to taking complex conjugation:
\be
X^{\mu}\rightarrow {X^{\mu}}^*, \quad
\Psi^{\a}\rightarrow {\Psi^{\a}}^*,
\eq
where $\ast$ denotes complex conjugation. In 
the following, we will use this to represent
string worldsheet parity transformation.

Thus we take $\G$ naturally to be a subgroup 
of $G_0\times\Z_2$, and any $g\in\G$ can be 
written as $(g_0, c(g))$, where $g_0\in G_0$ 
and $c(g)\in\Z_2$. The space we take quotient 
is thus actually the superspace times a factor 
of $\Z_2$ (or two copies of superspace), on 
which the worldsheet parity acts. The $Z_2$ 
structure has to be compatible with 
the multiplication in $G_0$:
\be
c(g_1 g_2)\equiv c(g_1)+c(g_2) \quad (\mbox{mod}\; 2),
\eq
where we used the representation with $c(g)=0,1$.
The quotient space $\R^d/\G$ is an orientifold if
$c(g)=1$ for some $g\in\G$.

The action of $g$ is
\bee
&X^{\mu}\rightarrow \Phi^{\mu}_g(X), \quad
\Psi^{\a}\rightarrow \Phi^{\a}_g(\Psi) \quad
\mbox{if}\; c(g)=0;
\label{trans1} \\
&X^{\mu}\rightarrow \Phi_g^{\mu}(X^{\ast}), \quad
\Psi^{\a}\rightarrow \Phi_g^{\a}(\Psi^{\ast}) \quad
\mbox{if}\; c(g)=1.
\label{trans2}
\eqq
In the above $\Phi_g$ is of the form
\be \label{P1}
\Phi_g^{\mu}(X)=R(g)^{\mu}_{\nu}X^{\nu}+a(g)^{\mu}, \quad
\Phi_g^{\a}(\Psi)=\Lam(g)^{\a}_{\b}\Psi^{\b},
\eq
where $(R(g),a(g))$ and $\Lam(g)$ are representations of $G_0$.
The map $\Phi_g$ does not depend on $c(g)$ but only on $g_0$.

Two consecutive transformations by $g_1$ and $g_2$ give
\be
X^{\mu}\rightarrow \Phi_{g_1}^{\mu}(X)
\rightarrow \Phi_{g_1}^{\mu}(\Phi_{g_2}(X)),
\eq
which should be equivalent to a single transformation by $g_1 g_2$:
\be
X^{\mu}\rightarrow \Phi^{\mu}_{g_1 g_2}(X).
\eq
It follows that
\bee
R(g_1 g_2)^{\mu}_{\nu}&=&R(g_1)^{\mu}_{\s}R(g_2)^{\s}_{\nu},
\label{Rgg} \\
a(g_1 g_2)^{\mu}&=&R(g_1)^{\mu}_{\nu}a(g_2)^{\nu}+a(g_1)^{\mu}.
\label{agg}
\eqq

Denote the operator of complex conjugation by $\Cc$
so that
\be \label{cc}
\Cc^2=1, \quad \mbox{and} \quad \Cc c=c^*\Cc
\eq
for any complex number $c$.
Eqs.(\ref{trans1}) and (\ref{trans2}) can now be summerized as
\be \label{transf}
X^{\mu}\rightarrow \Cc^{c(g)}\Phi^{\mu}_g(X)\Cc^{c(g)},
\quad
\Psi^{\a}\rightarrow \Cc^{c(g)}\Phi^{\a}_g(\Psi)\Cc^{c(g)},
\eq
where $\Cc^0=1$ and $\Cc^1=\Cc$.

Taking the quotient over $\G$ means that
we impose an equivalence relation between
$(X^{\mu}, \Psi^{\a})$ and their transformed images
(\ref{transf}).
Using this equivalence relation, we can divide the 
superspace into (possibly infinitely) 
many fundamental regions. Let there be $n$ partons 
in each fundamental region, and let each fundamental 
region be labelled by an element in the group $\G$.
We can first choose an arbitrary fundamental region
and label it by $e$, the unit element in $\G$.
Then the group element $g$ transforms the region
labelled by $e$ to another region labelled by $g$.
Hence the matrix variables acquire extra indices as
\be
X^{\mu}_{(g_1,i)(g_2,j)}, \quad \Psi^{\a}_{(g_1,i)(g_2,j)},
\eq
where $i,j=1,2,\cdots,n$ are the indices used
to label the $n$ partons in each fundamental region,
and $g_1, g_2\in\G$ are the indices used to label
the fundamental regions. In the flat covering space 
from which we take the quotient there are a total of 
$|\G|\times n$ partons. \footnote{By $|\G|$ we mean 
the number of elements in $\G$.} This implies that 
additional degrees of freedom appear upon compactification. 

With the notation that $\ast^0$ means the identity 
operation and $\ast^1$ means complex conjugation,
the equivalence relations are
\bee
X^{\mu}_{(gg_1,i)(gg_2,j)}&=&
R(g)^{\mu}_{\nu}{X^{\nu}}^{\ast^{c(g)}}_{(g_1,i)(g_2,j)}
+a(g)^{\mu}\d_{g_1 g_2}\d_{ij}, 
\label{qc0}\\
\Psi^{\a}_{(gg_1,i)(gg_2,j)}&=&
\Lam(g)^{\a}_{\b}\Psi_{(g_1,i)(g_2,j)}^{\b\ast^{c(g)}}.
\label{qc1}
\eqq
Note that in the D0-brane language the diagonal 
elements in the matrix variables are interpreted as 
the coordinates of the $n$ D0-branes,
and the off-diagonal elements represent open 
strings stretching between different D0-branes.
Thus if a translation is involved for a given $g$,
only the diagonal part of $X^{\mu}$ is shifted by 
the translation. This is why the factor  
$\d_{g_1 g_2}\d_{ij}$ appears in the last 
term of (\ref{qc0}).

We can rewrite the matrix equations (\ref{qc0}) 
and (\ref{qc1}) simply as
\bee
U(g)^{\dag}X^{\mu}U(g)&=&
\Cc^{c(g)}\Phi_g^{\mu}(X)\Cc^{c(g)}, \label{q1}\\
U(g)^{\dag}\Psi^{\a}U(g)&=&
\Cc^{c(g)}\Phi_g^{\a}(\Psi)\Cc^{c(g)} \label{q2}
\eqq
by using the matrices
\footnote{One can modify (\ref{U}) by a phase 
depending on $g$ and $g_1$, but it would be equivalent 
to (\ref{U}) by a unitary transformation.}
\be \label{U}
U(g)_{g_1 g_2}=\d_{g_1,gg_2},
\eq
which is nothing but a regular representation of $\G$.

Define
\be
\U(g)=U(g)\Cc^{c(g)}
\eq
then we have simpler expressions for (\ref{q1}), (\ref{q2}) as
\be \label{qc}
\U(g)^{\dag}X^{\mu}\U(g)=\Phi^{\mu}_g(X), \quad
\U(g)^{\dag}\Psi^{\a}\U(g)=\Phi^{\a}_g(\Psi).
\eq
In fact, eqs.(\ref{q1}), (\ref{q2}) can be understood
directly as a set of algebraic relations without 
going through the manipulation of indices in
the above arguments. Since the Matrix model is a 
gauge theory, eqs.(\ref{q1}), (\ref{q2}) mean that 
the equivalence relations are imposed as gauge 
transformations. We call eqs.(\ref{q1}) and (\ref{q2}), 
or equivalently (\ref{qc}), the quotient conditions.
It is important that we should treat the quotient 
conditions as algebraic relations. We are no longer 
restricted to define $U(g)$ by (\ref{U}), which only 
serves an illustrative purpose.

\section{Group Cohomology} \label{coho}

Without imposing (\ref{U}),
the compactification is not fully specified by
the quotient conditions,
although the geometric information is.
What we lack is the algebra of $\U$.
{}From (\ref{qc}) we see that
two successive transformations
of $X$ and $\Psi$ by $\U(g_1)$ and $\U(g_2)$
is equivalent to a single transformation
by $\U(g_1 g_2)$.
If we treat (\ref{qc}) as the only constraint
on $X$ and $\Psi$,
$\{\U(g)\}$ has to be a projective representation
of $\G$ \cite{GP,DM}:
\be \label{UU1}
\U(g_1)\U(g_2)=\U(g_1 g_2)e^{i\a(g_1,g_2)}.
\eq
In terms of $U(g)$ it is
\be \label{UU2}
U(g_1)U(g_2)^{\ast^{c(g_1)}}
=U(g_1 g_2)e^{i(-1)^{c(g_1 g_2)}\a(g_1,g_2)}.
\eq
If $\a(g_1,g_2)\equiv 0 \;\;(\mbox{mod}\; 2\pi)$
for all $g_1, g_2$, it is a genuine representation.
In general it corresponds to a twist of the equivalence
relation (\ref{qc0}), (\ref{qc1}) by a phase factor
determined by $\a$.

Since $\a(g_1,g_2)$ is a function
depending on two group elements
$(g_1,g_2)$ it is called a two-cochain.
The associativity of the product
\be
(\U(g_0)\U(g_1))\U(g_2)=\U(g_0)(\U(g_1)\U(g_2)),
\quad \forall g_0,g_1,g_2\in\G
\eq
implies that $\a$ has to be a two-cocycle,
that is, its coboundary vanishes
\be \label{ca}
\d\a\equiv 0 \quad (\mbox{mod}\; 2\pi).
\eq
The coboundary of $\a$ is defined as
\footnote{
The coboundary of a $k$-cochain is defined by
\be
(\d\om)(g_0,\cdots,g_k)=\om(g_1,\cdots,g_k)
+\sum_{l=1}^{k}(-1)^l\om(g_0,\cdots,g_{l-1}g_l,\cdots,g_k)
+(-1)^{k+1}\om(g_0,\cdots,g_{k-1})(-1)^{c(g_k)}.
\eq
}
\be
(\d\a)(g_0,g_1,g_2)=\a(g_1,g_2)-\a(g_0 g_1,g_2)
+\a(g_0,g_1 g_2)-\a(g_0,g_1)(-1)^{c(g_2)}.
\eq
Note that we have $\Z_2$-graded the usual definition
of coboundary operator for group cohomology.

To fully determine a compactification
we need to specify the two-cochain $\a$.
But not all different assignments of $\a$
result in physically different compactifications.
If we simply shift $\U(g)$ by a phase
\be
\U(g)\rightarrow \U(g)e^{i\b(g)},
\eq
the quotient conditions are not changed,
but $\a$ will be shifted by the coboundary of
the one-cochain $\b$ as
\be \label{ab}
\a\rightarrow \a+\d\b,
\eq
where
\be \label{cb}
(\d\b)(g_0,g_1)=\b(g_1)-\b(g_0 g_1)+\b(g_0)(-1)^{c(g_1)}.
\eq
Hence for a given compactification,
$\a$ is only defined up to the coboundary of a one-cochain,
i.e., $\a$ is a two-cocycle
defined in the second Hochschild cohomology $H^2(\G,U(1))$.

%Consider the transformation
%\be
%\Cc\rightarrow\Cc e^{i\om}.
%\eq
%It preserves the properties (\ref{cc}) of $\Cc$
%and results in the transformation
%\be
%\b(g)\rightarrow\b(g)+\d\om(g),
%\eq
%where
%\be \label{dom}
%\d\om(g)=\om-\om(-1)^{c(g)}
%\eq
%is the coboundary of the zero-cochain $\om$.
%Thus $\b$ is defined as an element in $H^1(\G,U(1))$.

We can always use the ambiguity (\ref{ab}) to choose
an $\a$ in a given equivalence class such that
\be \label{aa}
\a(e,g)=\a(g,e)=\a(g^{-1},g)=0,
\quad \forall g\in\G,
\eq
where $e$ is the unity in $\G$.
To do so, we first make a transformation (\ref{ab})
with $\b(e)=\a(e,e)$ so that $\a(e,g)=\a(g,e)=0$.
Then we can make another transformation
with $\b(g)=\a(g^{-1},g)/2$ in order to get $\a(g^{-1},g)=0$.
In the following we assume that (\ref{aa}) is satisfied,
so we can set
\be \label{Ue}
\U(e)=1 \quad \mbox{and} \quad \U(g^{-1})=\U(g)^{\dag}.
\eq

For the case of $\G=\Z_{2n}$ which is generated
by $e$ and $a$ with $a^{2n}=e$ and $c(a)=1$,
the second Hochschild cohomology is $H^2(\Z_{2n},U(1))=\Z_2$,
corresponding to the choice of sign in $\U(a)^{2n}=\pm 1$.
For the compactification on a two-torus,
the discrete group is $\Z^2$ and $H^2(\Z^2,U(1))$ is $U(1)$.
This phase factor was found to correspond to
the background three-form field $C$ on the two-torus
and the light-cone circle \cite{CDS,DH}.
In general, $H^2(\G,U(1))$ is the moduli space of certain
background parameters for a compactification.
For orientifolds not all possible choices of elements
in $H^2(\G,U(1))$ are physically acceptible
because some choices may lead to anomalous theories.
For instance, for the compactification of matrix model
on $S^1\times S^1/\Z_2$ to give the type I theory
\cite{BSS}, new fields have to be added for anomaly cancellation.
For the compactification on the orientifold $T^5/\Z_2$,
the anomaly-free condition eliminates certain choices of $\a$
\cite{FS,KR1}.

For the case of orbifolds there is a close analogy
with the motion of electrons in a background of
magnetic fields.
Because the wave function of an electron
changes by a phase $e^{i\int_{\g}dy^{\mu}A_{\mu}(y)}$
when moving along a path $\g$,
the translation operators $U(\g)$ do not commute, rather,
$U(\g_1)U(\g_2)=e^{i\th}U(\g_2)U(\g_1)$,
where $\th$ is given by the magnetic flux passing through
the closed region bounded by the sequence of movements:
$[\g_1,\g_2,-\g_1,-\g_2]$.
This interpretation suggests that the noncommutativity
between $U(g_1)$ and $U(g_2)$ in Matrix model compactifications
is related to the integration of a certain background field
over a closed region defined by the sequence of
actions by $g_1,g_2,g_1^{-1},g_2^{-1}$.
For the case of compactification on a two-torus
the background field is the two-form $B$-field
in string theory \cite{CDS,DH,CK},
or by duality the 3-form $C$ field in M theory.
The coboundary condition on $\a$ (\ref{ca}) means that
the field strength $H=dB$ vanishes.
So physically $\a$ represents the Wilson line degrees
of freedom for the $B$ field on two-cycles of
the compactified space.

\section{Solutions of Quotient Conditions} \label{Sols}

The quotient conditions can be solved by following
Zumino's prescription \cite{HWW}.
The idea is the same as ref.\cite{Tay}.
Since the state of any D0-brane or open string
can be related to those in the fundamental region
labelled by $e$, all the degrees of freedom reside in
the entries $X^{\mu}_{(g,i)(e,j)}$ and $\Psi^{\a}_{(g,i)(e,j)}$.
Let us denote them by $(A_g^{\mu})_{ij}$ and $(\Th_g^{\a})_{ij}$,
then we can use (\ref{qc}) to express $X$ and $\Psi$
in terms of $A$ and $\Th$.
These expressions are the solutions of quotient conditions.

Consider the Hilbert space
\be \label{Hilbert}
\H_0=\{ |g,i\ra: g\in\G, i=1,2,\cdots,n\}
\eq
with the inner product
\be
\la g_1,i|g_2,j\ra=\d_{g_1 g_2}\d_{ij}.
\eq
Then
\be
X^{\mu}_{(g_1,i)(g_2,j)}=\la g_1,i|X^{\mu}|g_2,j\ra
\eq
represents the open string stretching from 
the $i$-th D-particle in the fundamental region labelled by $g_1$
to the $j$-th D-particle in the fundamental region labelled by $g_2$
if $\la g_1,i|g_2,j\ra=0$.
If $\la g_1,i|g_2,j\ra=1$,
it represents the position of the $i$-th D-particle
in the fundamental region labelled by $g_1$.
Later we will see that $\H_0$ corresponds to
a trivial $U(n)$ bundle on the dual space.
We may choose other Hilbert spaces to represent
nontrivial bundles.

In the following we will omit the second part in $\H_0$,
that is, we will set $n=1$.
It is straightforward to put the factor of $|i\ra$ back later.
The state $|e\ra$ will be denoted ``$\ra$'' and
is called the vacuum.
The state $|g\ra$ can then also be defined by $U(g)\ra$.
The inner product on $\H_0$ implies that
\be \label{int-U}
\la U(g)\ra=\d_{ge}.
\eq
By shifting the phase of the vacuum we can make it real
\be \label{real-vac}
\ra^*=\;\ra.
\eq

Using the quotient condition (\ref{q1}), we find
\bee
X^{\mu}U(g)\ra&=&
U(g)\Phi_g^{\mu}(X^{\ast^{c(g)}})\ra \nn \\
&=&U(g)[R(g)^{\mu}_{\nu}{X^{\nu}}^{\ast^{c(g)}}+a(g)^{\mu}]\ra.
\eqq
Let
\be \label{XA}
X^{\mu}\ra=A^{\mu}(U)\ra,
\eq
where
\be
A^{\mu}(U)=\sum_{g}A_{g}^{\mu}U(g)
\eq
with $A_{g}^{\mu}=\Re(A^{\mu}_{g})+i\Im(A^{\mu}_{g})\in\C$.
Using (\ref{real-vac}) and (\ref{XA}), we find
\be
{X^{\nu}}^{\ast^{c(g)}}\ra=A^{\nu}(U)^{\ast^{c(g)}}\ra,
\eq
so
\bee
X^{\mu}U(g)\ra&=&U(g)\Phi^{\mu}_g(A(U)^{\ast^{c(g)}})\ra \nn\\
&=&\left[\sum_{g'}U(g')\Phi^{\mu}_{g'}(A(U)^{\ast^{c(g')}})
U(g')^{\dag}P(g')\right]U(g)\ra, \label{t}
\eqq
where we have used the projection operator $P(g')$ defined by
\be \label{proj}
P(g')U(g)\ra=\d_{g' g}U(g)\ra.
\eq
The last line in (\ref{t}) is an operator
independent of $g$ acting on $U(g)\ra$ for an arbitrary $g\in\G$.
It follows that
\be
X^{\mu}=\sum_{g}U(g)\Phi^{\mu}_{g}(A(U)^{\ast^{c(g)}})
U(g)^{\dag}P(g).
\eq

Using the identity
\be
P(g_1)P(g_2)=\d_{g_1 g_2}P(g_2),
\eq
which follows from (\ref{proj}),
we can express $X^{\mu}$ more concisely as
\be \label{sol-X}
X^{\mu}(A)=\At^{\nu}(\Ut)\Rt^{\mu}_{\nu}+\at^{\mu},
\eq
where
\bee
\Rt^{\mu}_{\nu}&=&\sum_{g}R(g)^{\mu}_{\nu}P(g), \label{dR} \\
\at^{\mu}&=&\sum_{g}a(g)^{\mu}P(g), \label{da} \\
\At^{\nu}(\Ut)&=&\sum_g\Ut(g)\At^{\nu}_g, \\
\Ut(g)&=&\sum_{g'}U(g')U(g)^{\ast^{c(g')}}U(g')^{\dag}P(g'),
\label{dU} \\
\At^{\nu}_g&=&\Re(A^{\nu}_g)+\ti\Im(A^{\nu}_g), \\
\ti&=&i\sum_g(-1)^{c(g)}P(g). \label{di}
\eqq
Eq.(\ref{sol-X}) is the most general solution
of $X$ on the Hilbert space $\H_0$.

Similarly, let
\be
\Psi^{\a}\ra=\Th^{\a}(U)\ra,
\eq
where
\be
\Th^{\a}(U)=\sum_g\Th^{\a}_g U(g),
\eq
then
\be
\Psi^{\a}=\tilde{\Th}^{\b}\tilde{\Lam}^{\a}_{\b},
\eq
where
\bee
\tilde{\Lam}^{\a}_{\b}&=&\sum_g\Lam(g)^{\a}_{\b}P(g), \\
\tilde{\Th}^{\a}(\Ut)&=&\sum_g\Ut(g)\tilde{\Th}^{\a}_g, \\
\tilde{\Th}^{\a}_g&=&\Re(\tilde{\Th}^{\a}_g)+
\ti\Im(\tilde{\Th}^{\a}_g).
\eqq
For $n>1$, $A^{\mu}_g$ and $\Th^{\a}_g$ become
$n\times n$ matrices which are constrained by
the Hermiticity of $X$ and $\Psi$.

To check directly that the quotient conditions (\ref{qc})
are satisfied by this solution,
we can use the following relations
\bee
\Rt^{\mu}_{\nu}U(g)&=&U(g)R(g)^{\mu}_{\s}\Rt^{\s}_{\nu}, \\
\at^{\mu}U(g)&=&U(g)(R(g)^{\mu}_{\nu}\at^{\nu}+a(g)^{\mu}), \\
\Ut(g)U(g')&=&U(g')\Ut(g)^{\ast^{c(g')}}, \\
\Ut(g)\U(g')&=&\U(g')\Ut(g), \\
\ti U(g)&=&U(g) \ti^{\ast^{c(g)}}, \\
\ti\U(g)&=&\U(g)\ti.
\eqq
To derive them we used (\ref{Rgg}),(\ref{agg}) and
\be
P(g)U(g')=U(g')P_{g'^{-1}g},
\eq
which follows from (\ref{proj}).

After the quotient conditions are solved,
the Matrix model is only concerned with $\Ut$, $\Rt$, $\at$
and $\ti$, which form a closed algebra:
\bee
&\Ut(g_1)\Ut(g_2)=e^{\ti\a(g_2,g_1)}\Ut(g_2 g_1), \label{Ut}\\
&[ \Rt^{\mu}_{\nu}, \at^{\s} ]=[ \Rt^{\mu}_{\nu}, \ti ]
=[ \at^{\mu}, \ti ]=0, \\
&\Rt^{\mu}_{\nu}\Ut(g)=\Ut(g)\Rt^{\mu}_{\s}R(g)^{\s}_{\nu}, \\
&\at^{\mu}\Ut(g)=\Ut(g)(\Rt^{\mu}_{\nu}a(g)^{\nu}+\at^{\mu}), \\
&\ti^2=1, \quad
\ti\Ut(g)=\Ut(g)\ti^{\ast^{c(g)}}. \label{last}
\eqq
Eqs.(\ref{Ut})-(\ref{last}) are derived
from (\ref{dR}),(\ref{da}),(\ref{dU}),(\ref{di}).
Now that we have obtained the algebra of $\Ut, \Rt, \at$
and $\ti$, we can forget about $U$.
Since $\Ut(g)\ra=U(g)\ra$, $\H_0$ can also be written as
$\{\Ut(g)\ra\}$.
The functional (\ref{int-U}) which determines the inner product
on $\H_0$ can be viewed as a functional on $\Ut$
\be \label{int-Ut}
\la\Ut(g)\ra=\d_{ge}.
\eq

Recall that in Sec.\ref{coho} we have assumed that
$U(g)$ is a matrix of numbers and thus
it commutes with any complex number (times the unit matrix).
In the solution of $X$, the operators $\Ut, \Rt, \at$
and $\At$ are also matrices of numbers,
so they all commute with the complex number $i$.
Thus $U(g)$ commutes with $i$
while $\U(g)$ commutes with $\ti$ for all $g\in\G$.
On the other hand, $\U(g)$ does not commute with $i$ if $c(g)=1$,
and $\Ut(g)$ does not commute with $\ti$ if $c(g)=1$.
In fact one can find another representation in which
the states are $\{\U(g)\ra\}$ where $\ti$ is represented
as $i$ times the unit matrix and $i$ is represented
by the nontrivial matrix which represents $\ti$
on the states $\{\Ut(g)\ra\}$.
Furthermore, the algebra of $\Ut$ is the same as
the algebra of right multiplication on the states
$\{\U(g)\ra\}$ by $\U$.
Roughly speaking, the algebra of $\U$ is dual to
the algebra of $\Ut$ and $i$ is dual to $\ti$.
This duality corresponds to the symmetry of
charge conjugation on D-branes.

The solution (\ref{sol-X}) of $X$ appears to be
of the same form as a covariant derivative on a quantum space.
The function $\At(\Ut)$ plays the role of gauge fields
and both $\at$ and $\Rt$ play the role of derivatives.
The algebra of $\Ut$ (\ref{Ut}) is viewed as
the algebra of functions on the base space of
the Matrix model as a gauge field theory.
For toroidal compactifications,
$X$ indeed turns out to be the usual covariant derivative
on the dual torus.
Note that the operator $\Rt$ is also a generalized derivative
on a quantum space.
For instance the covariant derivative
on the space of $\Z_n$ (discrete $n$ points)
is given by an operator of the form (\ref{sol-X})
without the last term $\at$.

\section{Noncommutative Geometry} \label{NC}

For completeness we briefly review in this section
elements of noncommutative geometry \cite{Con}.
The Riemannian structure of a quantum space is encoded in
the spectral triple $(\A, \H, \D)$ \cite{Con},
where $\A$ is the algebra of functions
on the quantum space,
$\H$ is the Hilbert space on which $\A$ is realized
as an operator algebra,
and $\D$ is the so-called ``Dirac operator''
which is a self-adjoint operator on $\H$.
The Dirac operator is used to define
the differential calculus on the quantum space.

The naive definition for integration on a quantum space
is just the trace over $\H$.
While $\H$ can be infinite dimensional,
the definition of integration may have to be regulated as
\be
\int f=\lim_{\Lam\rightarrow \infty}
\frac{\mbox{Tr}_{\H}(f e^{-|\D|^2/\Lam^2})}
{\mbox{Tr}_{\H}(e^{-|\D|^2/\Lam^2})},
\eq
which is reminicent of the regularization in quantum field theory.
The general definition of integration on a quantum space
is given by the Dixmier trace \cite{Con}.
The functional (\ref{int-Ut}) is generically different from
the integration for orientifolds although they always agree
for manifolds and orbifolds.
The most important property of integration is cyclicity,
that is,
\be
\int U^{\dag}f U=\int f
\eq
for a (reasonably well-behaved) unitary operator $U$ on $\H$.
The cyclity is important because it enables
us to find a gauge invariant action for a gauge theory
on a quantum space.

A differential $k$-form on a quantum space
is a formal expression like
\be \label{r}
\r=\sum_i a^i_0 da^i_1 \cdots da^i_k
\eq
where $a^i_j$, $j=0,1,\cdots,k$ are elements of $\A$.
An element of $\A$ can be multiplied to $\r$
from the left or from the right.
If it is multiplied from the right
one can use the Leibniz rule to rewrite
the product in the standard form (\ref{r}).
Denote the set of $k$-forms as $\Om^k\A$.
We define $\Om\A$ as $\oplus_k\Om^k\A$
on which the exterior derivative $d$ acts.
We impose on $\Om\A$ the nilpotency condition
\be
d^2=0
\eq
and the Leibniz rule
\be
d(\r_1\r_2)=(d\r_1)\r_2\pm\r_1(d\r_2),
\eq
where the sign depends on whether $\r_1$
is an even form ($+$) or an odd form ($-$).

For a given representation $\pi$ of $\A$ on $\H$
we can extend it to $\Om\A$ by defining
\be
\pi(\r)=\sum_i a^i_0 [ \D,a^i_1 ] \cdots [ \D,a^i_k ],
\eq
for a $k$-form expressed as in eq.(\ref{r}).
To simplify our notation here and below
we simply write $a$ to stand for $\pi(a)$.
The inner product of two differential $k$-forms
$\r_1, \r_2$ can be defined as
\be
\la\r_1|\r_2\ra=\int\pi(\r_1)^{\dag}\pi(\r_2).
\eq

The operator algebra on $\H$ induces
a differential calculus $\Om(\A)$ through
the representation $\pi$ by defining
\be \label{OA}
\Om(\A)=\oplus_k(\Om^k\A/J_k),
\eq
where
\be
J_k=\mbox{ker}\pi|_{\Om^k A}+d(\mbox{ker}\pi|_{\Om^{k-1}A}).
\eq
Therefore, two differential forms $\r_1,\r_2\in\Om^k\A$
are equivalent if $\pi(\r_1-\r_2)=0$
or if $(\r_1-\r_2)=d\r$ for a certain
differential form $\r\in\Om^{k-1}\A$ with $\pi(\r)=0$.
However, the representation $\pi$ does not
respect this equivalence relation.
We should therefore define another representation $\pih$
which is $\pi$ preceded by a projection onto
the subspace of $\Om^k\A$ orthogonal to $J_k$. 
\footnote{
Two differential forms are orthogonal if
their inner product vanishes.}

In our case of interest,
the Dirac operator is of the form
\be
\D=\g^{\mu}\D_{\mu},
\eq
where the $\g^{\mu}$'s are the usual gamma matrices
which commute with $\D$ and $\A$, and they satisfy
$\{\g^{\mu},\g^{\nu}\}=2\eta^{\mu\nu}$.
This type of Dirac operators were discussed in detail
in ref.\cite{HW1}, where it was shown that if the conditions
\bee
&[ \D_{\mu},\D_{\nu} ]=0, \label{DD}\\
&\pi(J_2)=\pi(\A) \label{J2}
\eqq
are satisfied, then
\be \label{drho}
\pih(d\r)=\frac{1}{2}\g^{\mu\nu}
([ \D_{\mu},\r_{\nu} ]-[ \D_{\nu},\r_{\mu} ])
\eq
for any differential one-form $\r$,
where $\pih(\r)=\g^{\mu}\r_{\mu}$ and
$\g^{\mu\nu}=\frac{1}{2}[\g^{\mu},\g^{\nu}]$.
The condition for (\ref{J2}) to be satisfied
is that any element in $\pi(\A)$ can be written as
\be \label{J22}
\sum_i a_i[ \eta^{\mu\nu}\D_{\mu}\D_{\nu}, b_i ]
\eq
for some $a_i, b_i\in\A$.

Let $A$ denote the gauge field in a Yang-Mills theory.
The gauge field strength is $F=dA+A^2$.
We define a dressed Dirac operator
\be
\Dt=D+\pih(A),
\eq
where $A$ is a one-form $\sum_i a_i db_i$
for some $a_i,b_i\in\A$ and
$\pih(A)=\sum_i a_i[ \D, b_i ]$.
Then it follows from (\ref{drho}) that
\be \label{F}
\pih(F)=\frac{1}{2}\g^{\mu\nu}[\Dt_{\mu},\Dt_{\nu}].
\eq
The action for a Yang-Mills theory on
a quantum space is defined to be \cite{Con}
\be
S_{YM}=\frac{1}{g_{YM}^2}\la F|F\ra.
\eq
The gauge field can be coupled to
a fermionic matter field $\Psi$
by adding the term \cite{HW1}
\be
S_f=\la\Psi|\Dt|\Psi\ra
\eq
and the total action is
\be \label{action}
S=S_{YM}+S_f.
\eq
It gives a super Yang-Mills theory (SYM) on a quantum space
for our matrix variables $X$ and $\Psi$ if
(\ref{F}) is satisfied \cite{HW1}.
If (\ref{F}) does not hold,
one can still define a Yang-Mills theory with matter
on a quantum space by the action (\ref{action}),
but then there will be new terms compared with
the standard SYM action.

\section{Quantum Space} \label{QS}

Interpreting the matrix model compactified on a quotient space
as a gauge field theory on a quantum space,
we need to specify the spectral triple $(\A,\H,\D)$
for the base space.
First, the algebra of functions $\A$ on the base space
is the algebra of $\Ut$
\footnote{A generic element in this algebra is of the form
$\sum_g a_g \Ut(g)$ with $a(g)\in\C$.}
because the gauge fields
$\At^{\mu}(\Ut)$ are functions of $\Ut$ for a trivial bundle.
We take the bare Dirac operator to be
\be
\D=\g_{\mu}X^{\mu}(A_0),
\eq
where $A_0^{\mu}$ are generic constant diagonal matrices.
This operator defines the calculus and
it has the property of flat connection
\be \label{flatconn}
[ X_{\mu}(A_0),X_{\nu}(A_0) ]=0
\eq
as in (\ref{DD}).

Given the solution $X^{\mu}(A_0)$ of the quotient conditions,
\be \label{X'}
X^{\mu}=\sum_i a_i(\Ut) X^{\mu}(A_0) b_i(\Ut)
\eq
is also a solution of the quotient condition (\ref{qc})
because $\Ut$ commutes with $\U$.
If $\sum_i a_i b_i=1$ then (\ref{X'}) can also be written as
\be
X^{\mu}=X^{\mu}(A_0)+\pih(A')^{\mu},
\eq
where $A'=\sum_i a_i db_i$ is a generic one-form field.
Thus we see that the interpretation of $X$ as
a covariant derivative, i.e, the partial derivative plus
a one-form field, is a very natural result of the fact
that $X$ is a solution of the quotient condition.
Conversely, we may take this as a guiding principle
for the study of gauge theories on noncommutative space.
We may even view the quotient condition as the fundamental
physical reason for the appearance of covariant derivatives
in gauge theories.

It follows that for generic $A_0$
the dressed Dirac operator is
\be
\Dt=\g_{\mu}X^{\mu}(A)
\eq
for an arbitrary function $A$ of $\Ut$.
Since $\H$ in the spectral triple is defined
to be the Hilbert space on which both $\A$ and $\D$
are realized as operators,
it should consist of both $\H_0$
and the space of Dirac spinors in $(9+1)$ dimensions.

Assuming (\ref{drho}), we find
\be
\pih(F)=\frac{1}{2}\g^{\mu\nu}F_{\mu\nu},
\eq
where
\be
F_{\mu\nu}=[ X_{\mu},X_{\nu} ].
\eq
Then (\ref{action}) gives precisely
the SYM action for the Matrix model.
The Matrix model compactified on the orientifold
of flat superspace over $\G$
is thus identified with the SYM theory on
the quantum space $(\A,\H,\D)$.

Define a set of one-forms by
\be
\xi_g=\Ut(g)^{\dag}d\Ut(g).
\eq
The Leibniz rule implies that
\be
\xi_{g'}\Ut(g)=\Ut(g)(\xi_{gg'}-\xi_g).
\eq
Using
\be
\pi(\xi_g)=\g_{\mu}\Rt^{\mu}_{\nu}
[(R(g)^{\nu}_{\s}-\d^{\nu}_{\s})A_0^{\s}+a(g)^{\nu}],
\eq
according to (\ref{OA}),
the differential calculus on the quantum space
satisfies the following commutation relations
\bee
&\{ \xi_g,\xi_{g'} \}=0.
\eqq
Then it also follows that
\be
\xi_g^2=d\xi_g=0.
\eq
In particular, if $R(g)^{\mu}_{\nu}=\d^{\mu}_{\nu}$
for a certain $g\in\G$, then
\be
[\xi_g,\Ut_{g'}]=0, \quad \forall g'\in\G.
\eq

In the above we have chosen a particular representation
for $\G$ which is the projective regular representation,
but we may choose other representations.
A different representation in general
means a different background.
If we take the Hilbert space $\H$ as above,
the gauge fields $A$ are defined as functions of $\Ut$'s,
which are functions on the quantum space.
This means that it lives on a trivial bundle.
The definition of a generic (twisted) bundle on a quantum space
is that the set of sections on a twisted bundle
is a projective module of $\A$ \cite{Con}.
\footnote{A projective module is a direct summand of
a (finitely generated) free module.}
This definition follows from the classical theorem that
all locally trivial finite dimensional complex vector bundles
over a compact space ${\cal M}$ are in one-to-one
correspondence with projective modules over the algebra of
continuous functions on ${\cal M}$.
For any projective module of $\A$ we can define
a right multiplication of $\A$ on it.
{}From our example of the regular representation,
we see that we should view the left multiplication
of $\U(g)$ as the right multiplication of $\Ut(g)$
for a generic projective representation,
while the left multiplication of $\Ut(g)$ is
in general not defined.
Sections on the bundle of adjoint representations
are represented by operators on the projective module
which commute with all $\U(g)$'s.
Indeed the space of solutions of $X$ to the quotient
conditions is the same space of operators on the
Hilbert space commuting with all $\U(g)$'s.

\section{Residual Gauge Symmetry} \label{ResSym}

Before the quotient conditions are imposed,
the gauge group is $U(N)$ where $N$ is the dimension
of the Hilbert space.
After the quotient conditions are imposed,
a gauge transformation by $\gh\in U(N)$
\be
X^{\mu}\rightarrow \gh^{\dag}X^{\mu}\gh
\eq
survives the quotient conditions (\ref{qc}) only if
\be \label{res}
\gh\U(h)\gh^{\dag}=\U(h)e^{i\b(h)},
\eq
or equivalently,
\be
\gh U(h)(\gh^{\dag})^{\ast^{c(h)}}=e^{i(-1)^{c(h)}\b(h)}U(h)
\eq
for a certain function $\b$ on $\G$.
Note that its consistency with the algebra of $U$
(\ref{UU2}) implies that the coboundary of $\b$ (\ref{cb})
\be \label{100}
\d\b(g_0,g_1)\equiv 0 \quad (\mbox{mod}\; 2\pi),
\eq
that is, $\b$ has to be a one-cocycle.
For $g_0=e$, eq.(\ref{100}) implies that $\b(e)=0$,
which is compatible with (\ref{Ue}).
Actually, $\b$ is only defined up to an exact one-cochain
because if we shift $\gh$ by a phase $e^{i\om}$,
$\b$ transforms as
\be
\b\rightarrow \b-\d\om,
\eq
where $\d\om(g)$ is given by
\be
\d\om(g)=\om-\om(-1)^{c(g)}.
\eq
Thus $\b$ is defined as an element in $H^1(\G,U(1))$.

According to (\ref{res}),
the transformation of $\Ut(h)$ by $\gh$ is
\be
\gh\Ut(h)\gh^{\dag}=e^{\ti(-1)^{c(h)}\b(h)}\Ut(h),
\eq
which induces a transformation on $A$ by 
\be
\At^{\mu}_h\rightarrow e^{\ti(-1)^{c(h)}\b(h)}\At^{\mu}_h.
\eq
For different Hilbert space $\H$ we have
different elements $\gh$ satisfying (\ref{res})
and thus different $\b$'s.

Obviously if $\gh$ is an arbitrary function of $\Ut$
then (\ref{res}) is satisfied with $\b=0$.
In fact these functions of $\Ut$ include all
the $\gh$'s with $\b=0$ for the Hilbert space $\H_0$,
otherwise (\ref{sol-X}) would not have been the most
general solution of $X$.
If $\gh$ is a function of $\Ut$ then it is
interpreted as a local gauge transformation on
the dual quantum space.
On the other hand, if $\b(g)\neq 0$ for some $g\in\G$,
$\gh$ indicates a global symmetry.
Note that the existence of $\gh$ depends on the choice of
the Hilbert space.

\section{Examples} \label{Example}

In this section we give a few examples to show
how the above formulation works for practical cases.
For completeness we also include examples of
manifold and orbifold as well as orientifold.
More examples can be found in \cite{HWW}.

\subsection{Torus}

A torus $T^d$ is a quotient space $\R^d/\Z^d$.
For simplicity consider a rectangular torus.
The action of the group $\G=\Z^d$ on $\R^d$ is
\be
\Phi^{i}_{\bf m}(X)=X^{i}+2\pi m_{i} R_{i}, \quad i=1,\cdots,d,
\eq
where ${\bf m}=(m_1,\cdots,m_d)$, $m_i\in\Z$,
is an element in $\G$.
This is a manifold so $c(g)=0$ for all $g\in\G$.
The algebra of $U$ is generically given by \cite{CDS}
\be
U({\bf m})U({\bf n})=e^{2i\a({\bf m},{\bf n})}U({\bf n})U({\bf m}),
\eq
where $\a({\bf m},{\bf n})=\th^{\mu\nu}m_{\mu}n_{\nu}$ and
$\th$ is an antisymmetric $d\times d$ matrix of real numbers.
The algebra of $\Ut$ is then
\be
\Ut({\bf m})\Ut({\bf n})=
e^{-2i\a({\bf m},{\bf n})}\Ut({\bf n})\Ut({\bf m}),
\eq
which is viewed as the algebra of functions
on the dual quantum torus.
A large amount of literature can be found on this subject (see,
for instance, refs.\cite{CDS,DH,CK,Ho,Cas,KO,AAS1,MZ,AAS2,BMZ}).

The solution of $X$ is that it is a covariant
derivative on the quantum torus.
We find $\Rt^{i}_{j}=\d^{i}_{j}$,
$\at^{i}=-i\del_{i}$ and $\ti=i$, where
$[\del_{i},\Ut({\bf m})]=im_{i}\Ut({\bf m})$
just like ordinary derivatives.
For trivial bundles
$X^{i}=-i\del_{i}+A^{i}(\Ut)$.

Integration over the quantum torus is given by
\be
\la\Ut({\bf m})\ra=\Pi_{i=1}^d\d_{m_i,0}
\eq
up to a normalization.
Eq.(\ref{F}) holds,
so the gauge invariant action (\ref{action}) is of
the same form as the SYM action on a classical torus.
The only new ingredient is that the base space
is not commutative.
The differential calculus is almost classical:
\be
[\Ut({\bf m}),\xi({\bf n})]=0.
\eq

Twisted bundles on the quantum torus are defined
to be projective modules of the algebra of $\Ut$.
Classically one can describe a twisted bundle
by twisted boundary conditions on sections of
the bundle.
For instance, for a section of the bundle in
the fundamental representation of $U(n)$,
we have the boundary conditions
\be
\phi(\s_1,\cdots,\s_j+2\pi,\cdots,\s_d)
=\Om_j\phi(\s_1,\cdots,\s_j,\cdots,\s_d)
\eq
for $j=1,\cdots,d$, and
$\Om$ is an $n\times n$ matrix of functions
of $(\s_1,\cdots,\s_{j-1},\s_{j+1},\cdots,\s_d)$
which specifies the gauge transformation of
the $n$-vector $\phi$ when $\s_j$ is shifted
by a whole cycle.
Note that the sections $\phi$ are in general not
functions well-defined on the torus, otherwise they
are sections of the trivial bundle.
Rather, the sections are functions on the infinite
plane with coordinates $\s_j$ viewed as
the covering space of the torus.

Here we make a general remark about twisted bundles
on quantum spaces.
In the classical case, sections of a twisted bundle
have the following properties.
First, one can multiply a function well-defined
on the torus from the right,
because this will not spoil the boundary condition.
Secondly, one can multiply a section of the adjoint
representation from the left for the same reason.
Thirdly, there is a definition of inner product
for two sections which gives a function
well-defined on the torus.
For the fundamental representation the inner product
of $\phi$ and $\phi'$ is just
$\sum_{i=1}^n\phi^{\dag}_i\phi'_i$.
In the quantum case, all three properties are preserved
The multiplication by functions of $\Ut$
from the right is defined to be the multiplication
by functions of $\U$ from the left,
which can always be defined since the $\U$'s are
always operators on the Hilbert space on which
the quotient conditions are defined.
The other two properties are preserved simply
because the twisted boundary conditions for the quantum
case are formally the same as for the classical case.
An explicit construction of the bundle by
imposing twisted boundary conditions on sections
is given in ref.\cite{Ho}.
Projective modules on quantum spaces or for
any $\G$ being a discrete Abelian group are
discussed in detail in ref.\cite{Rief}.

As mentioned in Sec.\ref{ResSym},
we need to impose gauge invariance conditions
on physical states for each element $\gh$
satisfying eq.(\ref{res}) with $\b=0$.
For the twisted bundle on the quantum torus $T_{\th}^2$,
sections of the adjoint representations are
generated by two sections $Z_1$ and $Z_2$,
both satisfying (\ref{res}) with $\b=0$.
Since \cite{CDS}
\be
Z_1 Z_2=e^{i2\pi\th'}Z_2 Z_1,
\eq
where $\th'=\frac{b-\th a}{n-\th m}$
and $a,b$ are integers satisfying $an-bm=1$,
they induce the transformations
\bee
&X_i=-i2\pi R_i D_i+A_i(Z_1,Z_2) \\
&\rightarrow\left\{
\begin{array}{l}
Z_1^{\dag}X_i Z_1=-i2\pi R_i D_i+A_i(Z_1,e^{i2\pi\th'}Z_2)
+\frac{2\pi R_i}{n-\th m}\d_{i1}, \\
Z_2^{\dag}X_i Z_2=-i2\pi R_i D_i+A_i(e^{-i2\pi\th'}Z_1,Z_2)
+\frac{2\pi R_i}{n-\th m}\d_{i2},
\end{array}\right.
\eqq
where $D_i$ is the covariant derivative on $T_{\th}^2$,
Let the operator conjugate to the constant part of $X_i$
be denoted $E_i$, and the operator shifting the
phase of $Z_i$ be denoted $P'_i$.
Then corresponding to the gauge invariance under
conjugations by $Z_i$,
the gauge transformation operator
\be
e^{i 2\pi\frac{R_i}{n-\th m}(E_i-\th'\eps_{ij}P'_j)}=\one
\eq
when acting on a physical state.
Since the phase of $Z_i$ is defined up to $2\pi$,
the eigenvalues of $P'_i$ are integers.
Let $P'_i=m_i\in\Z$, then the gauge invariance means
\be
E_i=\frac{n-\th m}{R_i}(n_i+\th'\eps_{ij}m_j)
\eq
for integer $n_i$.
This result differs from that in \cite{Ho},
for which proper corrections were pointed out in
\cite{HV2} and \cite{BM}.

\subsection{$A_{n-1}$} \label{Zn}

The ALE space $A_{n-1}$ is the quotient space $\R^4/\Z_n$.
The physics of D-branes on $A_{n-1}$ was
first described in ref.\cite{DM},
which in fact demonstrated the general principle
for D-branes on all kinds of quotient spaces.
Here we discuss this case as a simple example.

The action of $\G=\Z_n$ is generated by
\be
Z_i\rightarrow U^{\dag}Z_i U=\Phi_i(Z)=e^{i2\pi/n}Z_i,
\eq
where $Z_i=X_i+iX_{i+2}$ for $i=1,2$ and $m\in\Z_n$.
This is an orbifold so $c(m)=0$.
For a projective representation of $\Z_n$, 
$U^n=e^{i\th}$ for some $\th\in\R$.
By a change of phase of $U$
one can always set $\th=0$.
This means that $H^2(\G,U(1))$ is trivial
and we can set $\a=0$.
The algebra of functions on the quantum space is $\A=\Z_n$.

The bare Dirac operator on this quantum space is
$\Rt$ \cite{Con}, which satisfies
\be
\Rt U=e^{i2\pi/n}U\Rt.
\eq
Then a generic differential one-form $A$ has
$\pih(A)=f\Rt$ for some $f\in\A$.
It follows that the dressed Dirac operator
is also of the form $\Dt=f\Rt$.
Since $\at=0$, for a trivial bundle the solution of $Z$ is
\be
Z=A(U)\Rt,
\eq
which is understood as a covariant derivative on $\Z_n$.
Note that $\Ut=U$ in this case.

In the regular representation (for the trivial bundle),
$U$ and $\Rt$ are $n\times n$ matrices.
For instance we can choose
\be
U_{ij}=e^{i2j\pi/n}\d_{ij}, \quad
\Rt_{ij}=\d_{(i+1)j}.
\eq
Here $U$ is diagonal and it generates the algebra
of functions on $n$ points, which is represented
as a diagonal $n\times n$ matrix.
The commutator of $\Rt$ with a diagonal matrix
$f=\mbox{diag}(f_1,\cdots,f_n)$ is
$(f_{i+1}-f_i)\Rt_{ij}$,
with the factor of $(f_{i+1}-f_i)$
as one would intuitively expect for
a derivative for a discrete space.
Another choice of representation is
\be
U_{ij}=\d_{(i+1)j}, \quad
\Rt_{ij}=e^{-i2j\pi/n}\d_{ij}.
\eq
In this basis $U$ is off-diagonal,
corresponding to the interpretation that
it represents the open strings stretched
between the $n$ images of D-branes
in the covering space.
These two representations are in this sense dual
to each other.
The latter representation is more natural
in terms of D0-branes in the original compactified spacetime,
while the former representation is more natural
when considering the quantum base space $\Z_n$
on which the matrix model is viewed as a gauge theory.

It is easy to check that (\ref{drho})
is satisfied,
hence the action (\ref{action}) agrees with
the Matrix model action (SYM action).
The differential calculus is defined by
\be
U\xi=e^{-i2\pi/n}\xi U,
\eq
where $\xi=U^{\dag}dU$.
Projective modules on $\Z_n$ is discussed in ref.\cite{Rief}
and can also be obtained by the method given in ref.\cite{Ho}.

\subsection{Gauging World Sheet Parity}

Now we consider gauging only the world sheet parity.
The quotient condition for this case is
\be
U^{\dag}X^{\mu}U=X^{\mu\ast}.
\eq
The algebra of $U$ is defined by
\be
UU^{\ast}=\eps\one,
\eq
where $\eps=\pm 1$.
The choice of $\eps$ corresponds to the choice
of an element in $H^2(\Z_2,U(1))=\Z_2$.
A unitary transformation of $X$
\be
X\rightarrow V^{\dag}XV
\eq
is equivalent to a transformation on $U$ by
\be
U\rightarrow VUV^{T},
\eq
where $T$ denotes transposition.
It is straightforward to show that there are only
three inequivalent two-dimensional representations of $U$
\be
U=\left(\begin{array}{cc} 1 & 0 \\ 0 & 1 \end{array}\right),
\quad
  \left(\begin{array}{cc} 0 & 1 \\ 1 & 0 \end{array}\right),
\quad
  \left(\begin{array}{cc} 0 & -i \\ i & 0 \end{array}\right),
\eq
where the first two choices are for $\eps=1$
and the last choice is for $\eps=-1$.
It is amusing that although the first two cases are not
related by unitary transformations, they are equivalent
for the purpose of Matrix model, i.e.,
there exists a map between matrix variables for the two
choices so that they give the same action
for the Matrix model \cite{Pol}.
The reason why we are interested in two-dimensional
representations of $U$ is that we are interested in
projective regular representations of $\G=\Z_2$.
%It may seem strange at first sight that for
%a given choice of $\eps$ ($\eps=1$) there can be more
%than one inequivalent representations of $U$.
%However we note that the first possibility is not
%a regular representation.
%It is just the direct sum of two copies of
%the trivial representation $U=\one$.
%As it was illustrated in the example of
%the compactification on $T^4/\Z_2$ in ref.\cite{RW},
%such representations correspond to having
%D0-branes located at fixed points of the quotient space,
%and they may be interpreted as D-branes wrapped on
%vanishing cycles at the singularity.
%We will discuss more about this in the next section.
%For now we will focus on the other two possibilities,
%which are projective regular representations uniquely
%determined by our choice of $\eps$.

Using the solution (\ref{sol-X}) we find immediately that
\be
X=\left(\begin{array}{cc} A & B \\ B^{\dag} & D
\end{array}\right),
\eq
where $D^{\ast}=A$ and $B^{T}=\eps B$.

\subsection{$S^1\times S^1/\Z_2$}

In this example we consider the space $S^1\times S^1/\Z_2$
which is related to the heterotic matrix strings
\cite{DF,Mot,KR2,BSS,BMot,Rey,Low,KR,HWW}.

The quotient conditions are
\bee
U^{\dag}_i X_j U_i&=&X_j+2\pi\d_{ij}R_j, \quad i,j=1,2, \\
U^{\dag}_3 X_1 U_3&=&-X_1^{\ast}, \\
U^{\dag}_3 X_2 U_3&=&X_2^{\ast}.
\eqq
The algebra of $U$ is determined by a phase $q$
and two signs $\eps_1$, $\eps_2$ as \cite{HWW}
\bee
U_1 U_2&=&q U_2 U_1, \\
U_1 U_3&=&\eps_1 U_3 U_1^T, \\
U_2 U_3&=&U_3 U_2^{\ast}, \\
U_3 U_3^{\ast}&=&\eps_2\one.
\eqq
It follows that the algebra of $\Ut$ is
\bee
\Ut_2\Ut_1&=&\tilde{q}\Ut_1\Ut_2, \\
\Ut_3\Ut_1^{-1}&=&\eps_1\Ut_1\Ut_3, \\
\Ut_3\Ut_2&=&\Ut_2\Ut_3, \\
\Ut_3\Ut_3&=&\eps_2\one,
\eqq
where $\tilde{q}=e^{\ti h}$ for $q=e^{ih}$.
The case with $q=\eps_1=\eps_2=1$ was discussed
in the context of heterotic matrix string.
Hypermultiplets need to be added for the Matrix model
to be anomaly-free \cite{BSS,BMot}.
It would be interesting to see if other choices
lead to any consistent physical theory.

The solution of $X$ on a trivial bundle is
\be
X^i=\At^i(\Ut)\Rt+\at^i,
\eq
where $\Rt$ and $\at$'s are derivatives satisfying
\bee
\at_i\Ut_j&=&\Ut_i(\at_i+i\d_{ij}), \quad i,j=1,2,\\
\at_i\Ut_3&=&\Ut_3\at_i, \\
\Rt\Ut_i&=&\Ut_i\Rt, \\
\Rt\Ut_3&=&-\Ut_3\Rt.
\eqq

\section{Generalizations} \label{General}

In this section we point out several aspects
about matrix theory compactification that
are not fully developed in our formulation;
more work is needed for better understanding.

\subsection{More General Representations}

In previous sections we have assumed that the algebra of $\U$
is realized as a projective regular representation
of $\G$ on the $|\G|$ dimensional Hilbert space.
This corresponds to the assumption that there is no
D-brane located at a fixed point of the quotient space.
On the other hand, if there are D-branes located on
a fixed point in $\R^d/\G$, that is,
if the position of D-brane is invariant under the action
of a subgroup $H$ of $\G$,
then $U(h)$ should be realized by the unit matrix for all $h\in H$.
In ref.\cite{RW} it is nicely demonstrated for the case
of $T^4/\Z_2$ that such representations correspond to
the background of D2-branes wrapped on vanishing two-cycles
on the orbifold plane.
They have also considered direct sums of different representations.
In particular the regular representation of $\G$ can be
decomposed into a bunch of irreducible representations,
thus one can imagine the meeting of several D-branes at
various orbifold planes and merge into a D-particle which
can be separated from the orbifold plane \cite{RW}.
It is expected that their observation can be generalized
to other cases as well.

Consider a Hilbert space on which $U(h)=\one$ for all $h\in H$.
Then $U(h)$ is central in the algebra of $U$ so
$\a(h,g)=\a(g,h)=0$.
This means that certain choices of the $B$ field background
are not consistent with the background of D-branes located
at certain singularities.

\subsection{More General Quotient Conditions}

In the sense that the quotient conditions suggest
the more general algebra of $U$ for
the matrix compactification,
the formulation above also suggests some generalizations
of the Matrix model.

In the above we have mentioned that taking the quotient
of the flat space makes sense only if the map $\Phi$
is a symmetry of the Matrix model.
If we first compactify the Matrix model on
a torus and take the limit such that
the dual torus becomes an infinite space,
we get the gauge field theory on
a $d$-dimensional classical space or quantum space,
so that the matrix variables $X^{\mu}(\s)$
now depend on the coordinates $\s$ of the base space.
In the decompactification limit the symmetry
of Euclidean motions on the base space is restored,
so we can impose quotient conditions such as
\be
U(g)^{\dag}X^{\mu}(\s)U(g)=\Phi_g^{\mu}(X(\Lam_g(\s))),
\eq
where $\Lam_g(\s)$ is a Euclidean transformation of $\s$.
This kind of quotient conditions are considered
in ref.\cite{NS,Laz} and it is useful in describing
a resolution of the singularity in the quotient space.

\subsection{Noncommutative Spacetime to Begin With}

Another natural generalization is that since we are 
allowed to consider noncommutative space in the dual 
picture, we should also be allowed to consider Matrix 
models in which the entries of the matrix variables 
are noncommutative from the beginning.

For instance, let $X^i=X^i_a T^a$,
where $T^a$'s are the Lie algebra generators for $U(n)$
satisfying $[T^a,T^b]=if^{abc}T^c$.
We may consider the noncommutative space
\be \label{mn}
[X^1_a, X^2_b]=ih\d_{ab}.
\eq
Note that here the original space
is already a quantum plane even before
we compactify the Matrix model on a torus with
a background $B$ field.
The commutator of $X$ now becomes
\be
[X_1,X_2]=\frac{i}{2}\{X_{1a},X_{2b}\}f^{abc}T_c+ihc_2\one,
\eq
where $c_2$ is the second Casimir $\sum_a T^a T^a$.
Obviously this algebra is invariant under change
of basis of the Lie algebra.
This means that the subgroup $\Z_n$ of $U(n)$
is preserved and the $n$ D-branes are identical.
In addition the algebra is invariant under
rotations of $X_1$ and $X_2$,
although the global $SO(9,1)$ symmetry
is broken down to $SO(2)\times SO(7,1)$.
We can also compactify this model on torus as before
since the algebra is invariant under translation.
The only modification to our solution (\ref{sol-X})
is that now the coefficients $A_g^{\mu}$ are 
noncommutative.

Instead of the minimal noncommutativity (\ref{mn}),
one may even contemplate on other quantum spaces,
for instance those with quantum group symmetries.
While these are mathematically natural generalization
of the case with $h=0$, its physical interpretation
still needs to be identified.

%\subsection{Remarks}

%In general the matrix model action may have to be
%modified by terms such as the FI terms
%which is used to describe the blow-up of the 
%singularity \cite{Doug}. How can we naturally 
%incorporate them in the matrix compactifications?
%Another issue is whether higher loop effects are
%correctly reproduced by the Matrix model.
%In ref.\cite{GGR} a discrepancy is observed for
%the matrix model compactified on $T^4/\Z_2$.
%In general we are not sure what is the correct 
%action for matrix model on curved spaces,
%in particular on spaces with singularities.

%Recently in refs. \cite{HV1,HV2} Hofman 
%and Verlinde have suggested a Born-Infeld 
%Lagrangian on a noncommutative torus as a 
%compactified string theory. Part of their motivation
%is from an M-theory perspective. A natural question
%is: Can this theory be ``reproduced'' somehow
%in the Matrix model compactification?

%Another very important issue so far rarely discussed is 
%the quantization of gauge field theories on 
%quantum spaces, for instance on the quantum torus.

\section*{Acknowledgement}

P.M.H. thanks Prof. Chang-Yeong Lee for discussions 
and people at the Asia Pacific Center for Theoretical 
Physics for hospitality. His work is supported in part 
by the National Science Council, Taiwan, R.O.C. 
Y.S.W. thanks the warm hospitality of Laboratoire
de Physique Theorique, l'Ecole Normale Superieure 
in Paris. His work is supported in part by the U.S. 
National Science Foundation through grant PHY-9601277. 

%\appendix

%\end{references}
%\end{narrowtext}

\end{document}